\documentclass[preprint,tightenlines,aps,showpacs,showkeys,amsmath,amssymb]{revtex4}
\usepackage{graphicx}
\begin{document} 

\title{Excitation energy dependence of symmetry energy of
finite nuclei}

\author{S. K. Samaddar$^{1}$, J. N. De$^{1}$, X. Vi\~nas$^{2}$, and
M. Centelles$^{2}$}
\affiliation{
$^1$Saha Institute of Nuclear Physics, 1/AF Bidhannagar, Kolkata
{\sl 700064}, India \\
$^2$Departament d'Estructura i Constituents de la Mat\`eria,
Facultat de F\'{\i}sica, \\
and Institut de Ci\`encies del Cosmos, Universitat de Barcelona, \\
Diagonal {\sl 647}, {\sl 08028} Barcelona, Spain}


\begin{abstract}

A finite range density and momentum dependent effective interaction
is used to calculate the density and temperature dependence of the
symmetry energy coefficient $C_{sym}(\rho ,T)$ of infinite nuclear
matter. This symmetry energy is then used in the local density 
approximation to evaluate the excitation energy dependence of the
symmetry energy coefficient of finite nuclei in a microcanonical
formulation that accounts for thermal and expansion effects.
The results are in good harmony with the recently reported
experimental data from energetic nucleus-nucleus collisions.

\end{abstract} 



\maketitle

The symmetry energy $E_{sym} (\rho ,T)$ represents with a very good
accuracy the energy cost per nucleon to convert all the protons to
neutrons in symmetric nuclear matter at the density $\rho $ and
temperature $T$. Study of its density and energy dependence is of
utmost contemporary importance. It is essential not only for
understanding many aspects of exotic nuclear physics induced by
collisions of radioactive nuclei, but also a number of important
issues in the astrophysical scenario like supernovae explosions
\cite{bar}, explosive nucleosynthesis, cooling of protoneutron stars
\cite{lat} and abundances of relatively heavier elements. Even on a
more mundane level, the neutron skin thickness of heavier nuclei is
intimately correlated to the density derivative of the symmetry energy
\cite{bro,bal} as it reflects the pressure difference on the neutrons
and protons.

In addition to a kinetic contribution, the symmetry energy has a contribution
arising from the difference between the neutron-proton (n-p)
interaction and that between like pairs (n-n, p-p). Given an
interaction, it is straightforward to calculate the symmetry energy at
different densities and temperatures for infinite matter. There have
been several attempts in this direction. Calculations of the nuclear
equation of state (EoS) in the microscopic framework using both bare
\cite{dal,bal} and effective interactions \cite{bla,bar1} have been
done. The outcome of these calculations for the symmetry energy is
similar ($\sim 30-35$ MeV) at saturation density, but is considerably
different at subnormal as well as at supranormal densities where the
available data from experiment to confront with theory are more
scarce. 

Laboratory information on the density dependence of the symmetry energy
can be obtained from energetic nucleus-nucleus collision experiments.
At densities above the normal, it can be inferred from the comparison
of theoretical predictions with experimental data on the differential
flow of neutrons and protons, from the $\pi^- /\pi^+$, $K^0 /K^+ $
ratios, etc \cite{chen}. At subnormal densities, disassembly of a hot
expanded nucleus offers the best tool to study the characteristics of
the symmetry energy \cite{sou,she}. Experimental data related to
isotopic distributions, isospin diffusion, and isoscaling try to
constrain the density dependence in the subnormal region, but there is
still considerable uncertainty.

A nucleus expands with excitation with increasing temperature in
general. This implies an excitation energy dependence of the symmetry
energy because of the density change. Experimentally, this information
is generally extracted \cite{she} from the fit of the experimental
isotopic distributions at different excitation energies to those
obtained from a model for multifragmentation like the statistical
multifragmentation model (SMM) \cite{bon} or from isoscaling
\cite{ono}. Currently, calculations for the energy dependence of the
symmetry energy are available for infinite matter, but no microscopic
calculation has yet been performed for the energy dependence of finite
nuclei. The main purpose of this communication is to report such a
calculation. 

 For an expanding system pursuing the equilibrium configuration,
the surface diffuseness is likely to play an important role \cite{sob1},
thus a zero-range interaction like the Skyrme force widely used
to explore the nuclear ground-state properties may not be the most adequate
for generating such a density profile. It is further noted that
a constrained expanded system in a Thomas-Fermi approach may lead
to numerical instabilities \cite{lom} and the gradient (surface)
terms in the energy density functional were replaced with a suitable
Yukawa interaction \cite{dav}.
We have therefore chosen the modified Seyler-Blanchard (SBM) effective
interaction \cite{ban} for our microscopic calculation in the
finite temperature Thomas-Fermi formulation. This interaction
is of finite range and momentum and density dependent. The
interaction reproduces quite satisfactorily the ground state
bulk properties of nuclei over the whole periodic table for
$A>16$. The EoS calculated \cite{uma} with this interaction  agrees
very favorably with those obtained microscopically with a realistic
interaction in a variational approach \cite{fri,wir}. The SBM
interaction is given by
\begin{eqnarray}
v({\bf r}_1,{\bf r}_2,p,\rho)= -C_{l,u}\left [1-
\frac{p^2}{b^2}-d^2\left \{\rho ({\bf r}_1)+\rho ({\bf r}_2)\right
\}^n \right ] \frac{\exp(-r/a)}{(r/a)}.
\end{eqnarray}
 Here $r= |{\bf r}_1-{\bf r}_2|$ and $p= |{\bf p}_1-{\bf p}_2|$ are
the relative separation of the interacting nucleons in coordinate
and momentum space, $\rho ({\bf r}_1)$ and $\rho ({\bf r}_2)$
are the densities at the sites of the two interacting nucleons, and $C_l$
and $C_u$ are the strengths for like pair and unlike
pair nucleon-nucleon interaction. The density exponent $n$ controls
the stiffness of the nuclear EoS. The values of the parameters
$C_l$, $C_u$, $b$, $d$, $a$ and $n$ are given in Refs. \cite{ban,de}.
The energy per nucleon $e(\rho ,T)$ calculated with this interaction
for infinite nuclear matter is given by
\begin{eqnarray}
e(\rho ,T)=\frac{1}{\rho}\sum_\tau \rho_\tau \left[TJ_{3/2}(\eta_\tau )/
J_{1/2}(\eta_\tau )\left\{1-m_\tau^kV_\tau^1\right\}+\frac{1}{2}
V_\tau^0 \right ],
\end{eqnarray}
where $\tau$ refers to the isospin index (n, p). Here $J_q(\eta )$
are the Fermi integrals, $m_\tau^k$ the effective $k-$mass of
the nucleon and $\eta_\tau$ the fugacity given by
\begin{eqnarray}
\eta_\tau =(\mu_\tau -V_\tau^0 -V_\tau^2)/T,
\end{eqnarray}
with $\mu_\tau$ as the nucleon chemical potential. In Eqs.~(2) and (3),
the $V_\tau$'s are the different components of the single-particle
potential whose expressions can be found in \cite{ban}.

The symmetry energy per nucleon of asymmetric nuclear matter with asymmetry
$X=(\rho_n -\rho_p)/\rho$ is
\begin{eqnarray}
e_{sym}(\rho ,T,X)=e(\rho ,T,X)-e(\rho ,T,X=0).
\end{eqnarray}
It can be written as
\begin{eqnarray}
e_{sym}(\rho ,T,X)= C_{sym}(\rho ,T)X^2 + {\cal O}(X^4).
\end{eqnarray}
The terms beyond $X^2$ are negligible over a considerable range
of $X$ (as involved in finite nuclei). The symmetry energy coefficient
$C_{sym}$ is obtained from
\begin{eqnarray}
C_{sym}(\rho ,T)=\frac{1}{2}\frac{\partial^2}{\partial X^2}
e_{sym}(\rho ,T,X)|_{X=0}.
\end{eqnarray}

In the top panel of Fig.~1, the density dependence  
of $C_{sym}$  at $T=0$ is displayed in the density region
$0.1 \leq \rho /\rho_0 \leq 1.0$ for three variants of the
SBM interaction with density exponents $n=1/6$, 2/3, and 4/3,
in increasing order of the stiffness of the nuclear EoS. The
values of the nuclear incompressibility with these three interactions
are $K_{\infty} =238$, 300, and 380 MeV. The
symmetry energy coefficient calculated with them
can be very nicely represented
by $C_{sym}(\rho ) \sim C_{sym}(\rho_0) \, (\rho /
\rho_0)^{\gamma }$ with $C_{sym}(\rho_0)=34.0$ MeV and
$\gamma $= 0.65, 0.68, and 0.70, respectively. The value of
the exponent $\gamma $ appears not very sensitive to the
nuclear EoS. The agreement of the functional form of the symmetry
energy coefficient  and the value of $\gamma $ with those obtained
recently \cite{she} from experimental data 
($\gamma \simeq$ 0.69) is excellent. 
The nuclear incompressibility with $n=1/6$ compares very
well with the presently accepted value of $K_{\infty} \sim 
230$ MeV; all the subsequent calculations are therefore reported
for the SBM interaction with $n=1/6$. In the
bottom panel of Fig.~1, the temperature dependence of
the symmetry energy coefficient at different densities ($\rho /\rho_0 
=0.3$, 0.6, and 1.0) is shown. At a fixed density, dependence on
temperature is not much evident.  

 In calculating the excitation energy or density dependence of 
the symmetry energy coefficient of a finite nucleus with $N$
neutrons and $Z$ protons ($A=N+Z$) with excitation energy
$E^*$, we remind ourselves that the hot nucleus 
prepared in the laboratory in energetic nuclear collisions
is an isolated system with a fixed total 
excitation $E^*$ and thus should be described by microcanonical
thermodynamics \cite{sob}. Left to itself, the 
system expands due to unbalanced
thermal pressure in search of maximal entropy where the total
pressure vanishes and the system is in equilibrium in a bloated
mononuclear configuration with the same excitation energy.
The expansion is simulated through a self-similar scaling 
approximation for the density:
\begin{eqnarray}
\rho_\lambda (r)=\lambda^3\rho (\lambda r),
\end{eqnarray}
where $\lambda $ is the scaling parameter ($0< \lambda \leq 1$) and
$\rho (r)$ is the base density profile. The base density, employing
the SBM interaction, is generated in the self-consistent Thomas-Fermi
framework. The subtraction scheme \cite{bonc,sur} is used
to render the density profile independent of the box size with
an effective temperature $T$ chosen so as to give the maximum
entropy for the given excitation $E^*$. The excitation energy
is calculated as  
\begin{eqnarray}
E^*~=E(\lambda_{eq},T)-E(\lambda =1,T=0),
\end{eqnarray}
where $\lambda_{eq}$ is the scaling parameter for the equilibrium
density profile corresponding to this excitation. 

The SBM interaction,
being momentum dependent, renormalizes the bare nucleon mass
$m$ to an effective $k-$mass. A frequency dependent mass factor
$m_{\omega}/m$ is further phenomenologically incorporated
\cite{pra,shl} in the calculation. It is very relevant in the
present context; the $\omega $-mass $m_{\omega}/m$ is generally
larger than unity, it has the effect of bringing down the 
excited states from higher to lower energy near the 
Fermi surface, thus increasing the many-body density of states at
low excitations that allow comparatively more accommodation of entropy
at a given excitation energy. Details on the generation of the  
equilibrium density profile, effective temperature, frequency dependent
mass, etc., as employed in this calculation,
are given in Refs. \cite{sam,de1}.

Once the equilibrium density $\rho (r)$ of a nucleus at excitation
$E^*$ is known, the symmetry energy is calculated in the local density
approximation as
\begin{eqnarray}
C_{sym}(E^*)\left (\frac {N-Z}{A}\right )^2=\frac {1}{A}\int \rho (r)
\, C_{sym}^l(\rho (r),T)\left (\frac {\rho_n(r)-\rho_p(r)}
{\rho (r)}\right )^2 d{\bf r}.
\end{eqnarray}
Here $C_{sym}^l(\rho (r),T)$ is the symmetry energy coefficient at
temperature $T$ of infinite nuclear matter at a value of the local
density $\rho (r)$. The local isospin density is given by
$\rho_n(r)-\rho_p(r)$. It may be mentioned that both the volume
and the surface terms in the liquid drop type mass formula are
asymmetry dependent \cite{mye}. 
The symmetry energy coefficient $C_{sym}(E^*)$
defined through Eq.~(9) may therefore be taken as an effective
parameter incorporating both the volume and surface contributions
from asymmetry and may be written as
\begin{eqnarray}
C_{sym}=C_{sym}^{vol}-C_{sym}^{surf}/A^{1/3}.
\end{eqnarray}

In a microcanonical formulation, it has been found that
the equilibrium density at a given excitation depends on the mass
and asymmetry of the nucleus concerned \cite{sam}. In investigating
the excitation energy dependence of the symmetry energy coefficient,
it would then be worthwhile to investigate its system dependence.
We have therefore chosen three systems, two isobars of $A=150$,
namely, Cs and  Sm,  and a lighter system $^{40}$S. In the bottom
panel of Fig.~2, the coefficient $C_{sym}$ as calculated using Eq.~(9)
is displayed as a function of $E^*/A$ for the isobars of
$A=150$. It is found that at a fixed excitation, including the
ground state, $C_{sym}$ is somewhat sensitive to the asymmetry of the
nucleus; it increases with increasing proton fraction of the system. 
This is at variance with the expectation from the
liquid-drop formula where the effective symmetry energy coefficient
given by Eq.~(10) is independent of charge for a given mass.
This may be understood from the fact that the parameters 
of the liquid-drop formula are based on a global fit to the
binding energies of nuclei over the entire periodic table
around the stability line
excluding the very light ones. Here, the Coulomb energy
($=a_cZ^2/A^{1/3}$) coefficient $a_c$ is taken as a constant
for the whole mass range. In our calculations for isobars, it
is seen that with increasing charge as the proton distribution
is pushed outward, $a_c$ decreases and $C_{sym}$ increases. This
effectively explains the isobaric variation of binding energies.

Some representative experimental results \cite{she} for $C_{sym}$
obtained from the analysis of isoscaling data for lighter fragments
are shown in the bottom panel of Fig.~2 as inverted open triangles and
solid circles. At relatively lower excitations ($E^*/A\sim$ 2-3 MeV),
experimental  data from isoscaling for heavier fragments \cite{sou1}
are also shown as solid squares.
Our calculated energy dependence of the symmetry energy
coefficient agrees favorably with these experimental findings,
the calculated values are somewhat lower. The
variation of $C_{sym}$ with excitation stems basically from the
changing equilibrium density with excitation energy, which is
shown in the middle panel of the figure for the two systems. Since
the density has a profile, the choice of a single value of the 
density leaves room for ambiguity; we have taken $\rho_c/\rho_{c,0} $
as the measure of the density where $\rho_c$ is the central equilibrium
density at the relevant excitation and $\rho_{c,0}$ is the ground
state central density. The theoretically calculated results are in
nice agreement with the experimental data from Ref. \cite{nat} (solid
circles of the middle panel) derived from the analysis of caloric
curve measurements. The data obtained from Coulomb barrier systematics
\cite{brack,vio} are shown with open squares. For completeness, in the
top panel, the caloric curves along with the experimental data
compiled by Cibor {\it et al} \cite{cib} are also displayed. It is
seen that the plateau of the caloric curve shows little sensitivity
to the asymmetry of the nucleus; this is consistent with the 
recent calculations of Hoel {\it et al} \cite{hoe}.

The mass dependence of the effective symmetry energy $C_{sym}(E^*/A)$
is displayed in the bottom panel of Fig.~3. The calculations are done
for $^{150}$Sm and $^{40}$S, both having nearly the same asymmetry.
The reduction in $C_{sym}$ for $^{40}$S can be understood from the 
role played by the surface asymmetry as given by Eq.~(10). For
completeness, the equilibrium central densities and temperatures
as a function of $E^*/A$ are also shown in the middle and top panel
of the figure, respectively.

To conclude, calculations on the density and excitation
energy dependence of the symmetry energy of finite nuclei
have been reported in this communication in a microscopic
formulation within the microcanonical framework. It has been
stressed in a recent calculation \cite{sob1} that the surface 
diffuseness of the expanded mononuclear system plays a key
role in making the system softer towards instability, limiting
the maximum excitation energy a mononucleus can hold to $\sim 5$
MeV/A with free variation of the surface diffuseness. 
Our model calculation does not leave any room for free
variation of the surface diffuseness. 
It is determined in two stages: the increased diffuseness of
the base density profile of the hot nucleus 
over that of the ground state and then
its subsequent stretching  from the self-similar expansion. The
surface diffuseness so obtained is found to be somewhat less than
that reported in \cite{sob1,hoe}. Exploring the weakening of the
symmetry energy with excitation using free variation of the surface 
diffuseness would be interesting to look into.   
In the present calculation, the density dependence of $C_{sym}(\rho )$
of infinite nuclear matter is found out to be $\sim (\rho /\rho_0)
^\gamma$ with $\gamma =0.65$, very close to the recently extracted
experimental value of $\gamma \simeq 0.69$ \cite{she}. At constant
density, $C_{sym}(E^*)$ of infinite nuclear matter is practically
constant. For finite nuclei, however, density changes with
excitation; their
excitation energy dependence can be well represented by
$C_{sym}(E^*) \simeq C_{sym}(E^*=0)(1-\alpha E^*)$
with $\alpha \simeq 0.06$. These are in good consonance with the
experimental data obtained from nuclear multifragmentation.

\acknowledgments{S.K.S. and J.N.D. acknowledge the financial support
from CSIR and DST, Government of India, respectively. M.C. and X.V.
acknowledge financial support from Grants No.\ FIS2005-03142 from MEC
(Spain) and FEDER, and No.\ 2005SGR-00343 from Generalitat de
Catalunya.}

\newpage

\centerline
{\bf Figure Captions}
\begin{itemize}
\item[Fig.\ 1] Symmetry energy coefficient of infinite nuclear
matter as a function of density 
for different variants of the SBM interaction (see text) at
$T=0.0$ MeV (top panel). In the bottom panel the temperature
dependence of $C_{sym}$ at several fixed densities is
shown for $n=1/6$.

\item[Fig.\ 2] The equilibrium temperature (top panel), equilibrium
central density (middle panel) and the symmetry energy coefficient
(bottom panel) as a function of excitation energy for the $A=150$
isobars (Cs and  Sm). 
The experimental data for $T$ are from Ref.\ \cite{cib}, those for
$\rho/\rho_0$ are from Refs.\ \cite{nat} (circles) and
\cite{brack,vio} (squares), and those for $C_{sym}$ are from Refs.\
\cite{she} (inverted triangles and circles) and \cite{sou1} (squares).

\item[Fig.\ 3] Same as in Fig.~2 for the systems $^{40}$S and
$^{150}$Sm to show the mass dependence.

\end{itemize}

\newpage

\begin{figure}[t]
\includegraphics[width=0.75\textwidth,angle=0,clip=false]{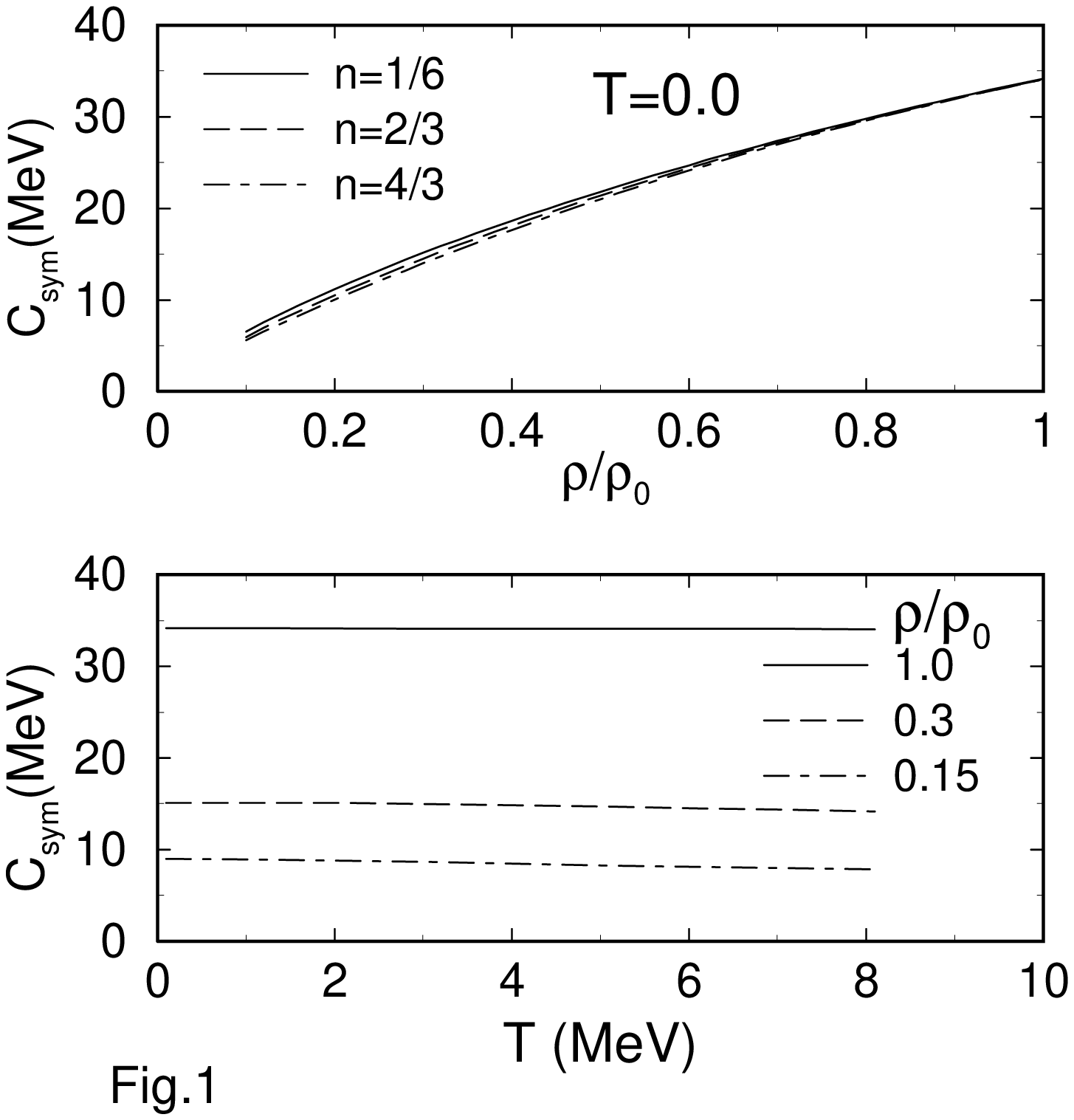}
\end{figure}

\begin{figure}[t]
\includegraphics[width=0.75\textwidth,angle=0,clip=false]{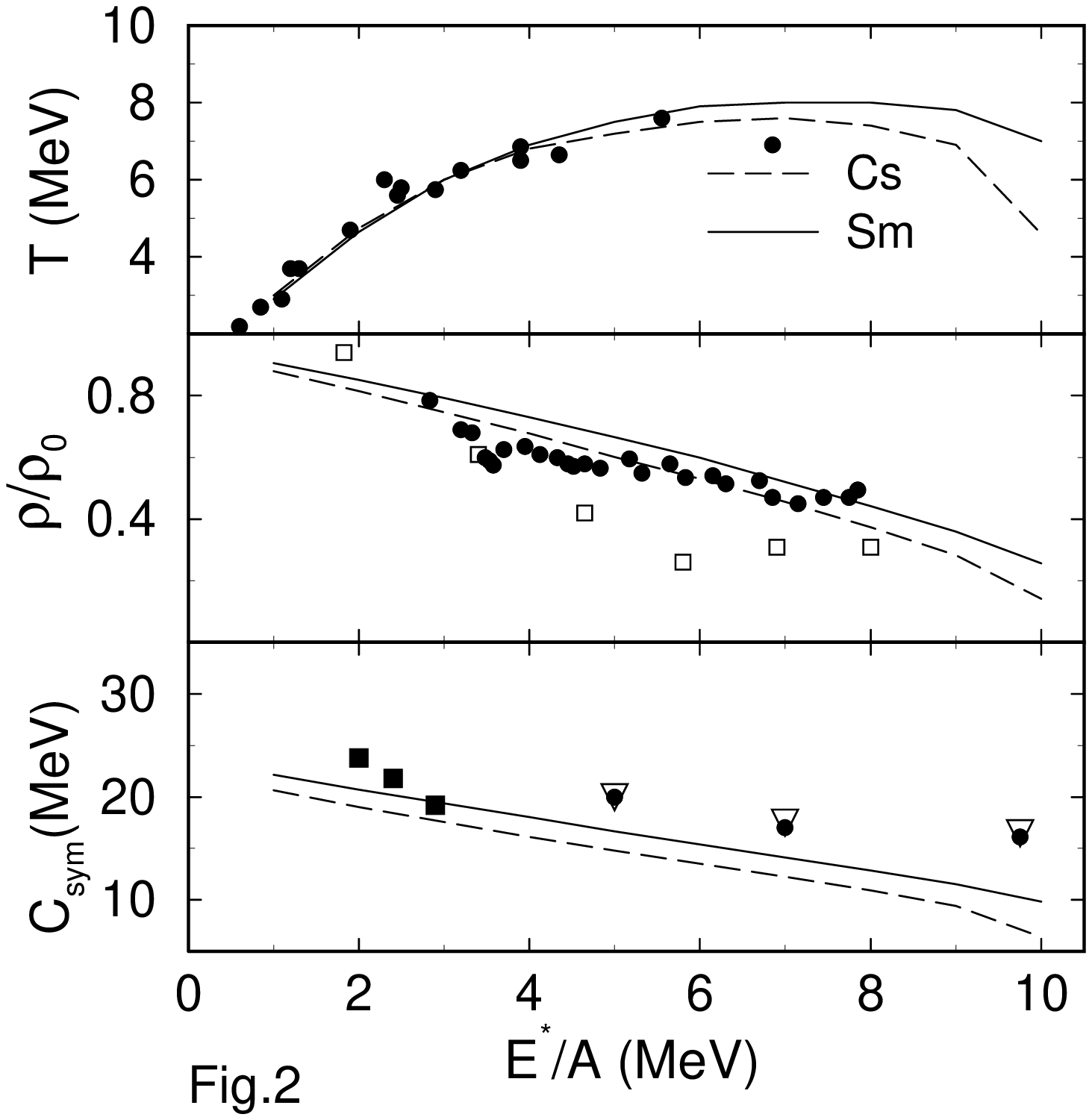}
\end{figure}

\begin{figure}[t]
\includegraphics[width=0.75\textwidth,angle=0,clip=false]{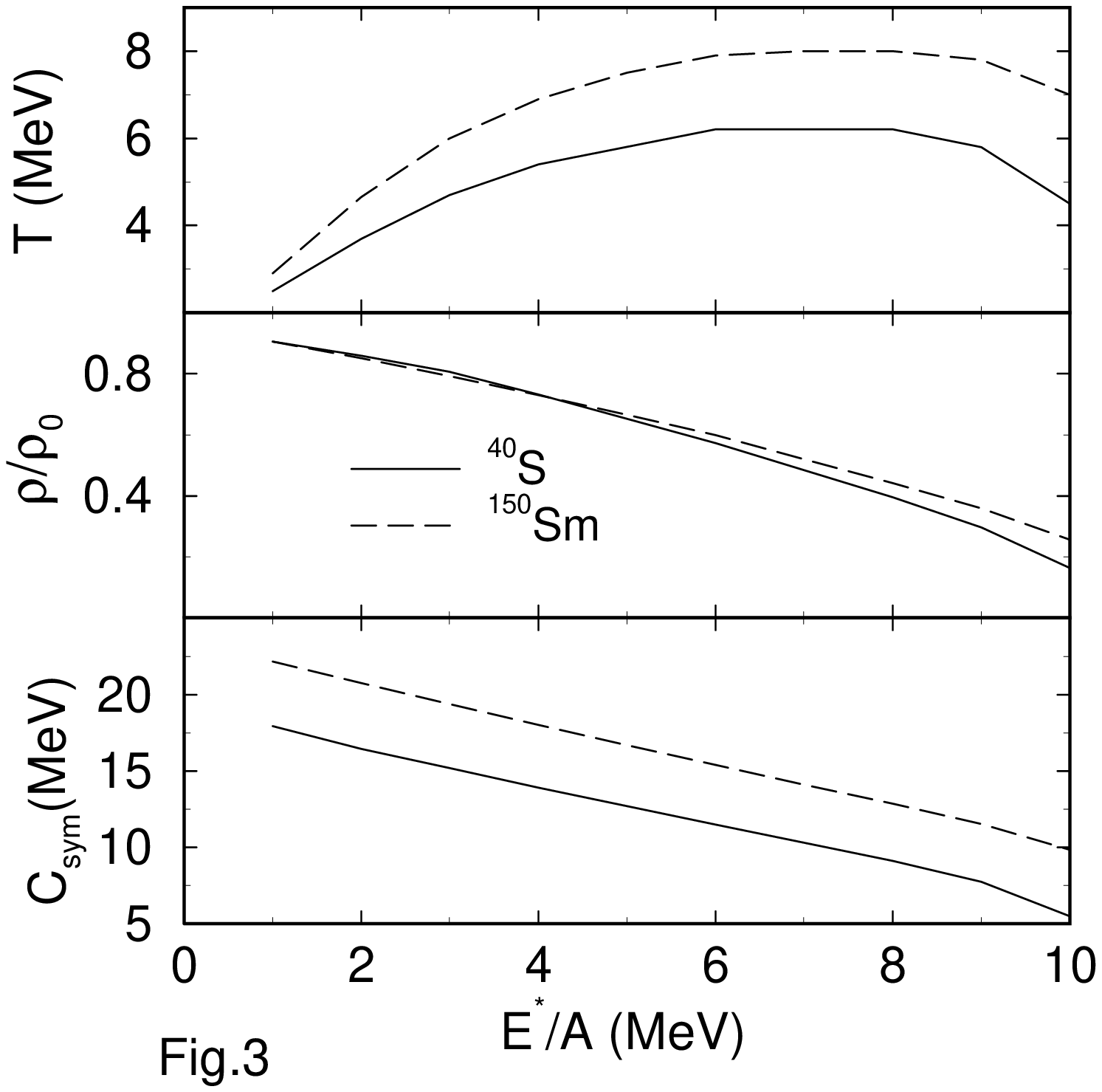}
\end{figure}

\end{document}